\documentclass[preprintnumbers,amsmath,amssymb,floatfix,10pt,prd,onecolumn,
superscriptaddress,nofootinbib]{revtex4}

\usepackage{amsfonts}
\usepackage{mathrsfs}
\usepackage{latexsym}
\usepackage{epsfig}
\usepackage{epstopdf}
\usepackage{graphicx}
\usepackage{amssymb}
\usepackage{amsmath}
\usepackage{dcolumn}
\usepackage{bm}
\usepackage{color}
\usepackage{xcolor}
\usepackage{comment}
\usepackage[cp1251]{inputenc}
\begin{document}

\title{\bf Anisotropic Quark Stars in Modified $f(R,T)$ Gravity utilizing Tolman V potential}

\author{Tayyaba Naz}
\email{tayyaba.naz@nu.edu.pk}\affiliation{National University of Computer and Emerging Sciences,\\ Lahore Campus, Pakistan.}

\author{Adnan Malik}
\email{adnan.malik@zjnu.edu.cn; adnanmalik_chheena@yahoo.com; adnan.malik@skt.umt.edu.pk}
\affiliation{School of Mathematical Sciences, Zhejiang Normal University, \\Jinhua, Zhejiang, China.}
\affiliation{Department of Mathematics, University of Management and Technology,\\ Sialkot Campus, Lahore, Pakistan}

\author{Zenab Ramay}
\email{zenabramay88@gmail.com}\affiliation{National University of Computer and Emerging Sciences,\\ Lahore Campus, Pakistan.}

\begin{abstract}
\begin{center}
\textbf{Abstract}\\
\end{center}
Alternative gravity theory is currently an incredibly significant technique for addressing some enduring experimental difficulties, such as the universe's dark region. They may also be employed in celestial cosmology, producing results that are a stage beyond those found using Einstein's General Relativity. In this study, we examine the characteristics of anisotropic spherically symmetric stellar structures in the context of modified $f(R,T)$ gravity. In order to explain the distinctive characteristics of compact objects, we investigate how the fluid distribution in the star model is affected by the MIT bag model equation of state. By using Tolman V metric potentials, we establish the field equations and by employing the experimental data of observed three stars, we identify the values of unknown parameters. By using realistic $f(R,T)$ model, we investigate the effect of the energy density, anisotropic factor, transversal and radial pressure within the cores of the aforementioned stars for a particular amount of the Bag constant. Further, we examine the stability of the cosmic structure and the physical validity of our suggested model via equilibrium conditions, energy and causality parameters. To conclude, the physical conditions are fulfilled by our model, and the magnitude of the Bag constant agrees with the experimental data, demonstrating the model's feasibility.
\end{abstract}
\maketitle

\section{Introduction}
Despite being nearly a century old, Einstein's general relativity theory $(GR)$, which was suggested in 1915, is the most accurate modern physics theory used for explaining gravitational aspect. Indeed, $GR$ predictions have been successful in all experimental evaluations but numerous challenges cannot be tackled by $GR$, both theoretically and empirically \cite{Coley}. Most relativistic astrophysicists have been motivated to alter $GR$ as a result of latest astrophysical data concerning the dark matter $(DM)$ problem and the speeding up of cosmic expansion. It is stated that a mysterious energy source known as dark energy $(DE)$ that has unknown characteristics is responsible for this universal expansion. Alternate theories of $GR$ have been crucial in revealing the secrets of $DE$ and $DM$. In the discussion of the features of modified relativistic behaviour, Qadir et al. \cite{Qadir} came to the conclusion that $GR$ might have to be changed as a potential solution to some cosmological problems, such as quantum gravity and the $DM$ issues. In this framework, various methods have been taken into consideration to tackle these challenges. \\
Researchers have proposed altered theories of gravity (that broaden Einstein's theory of $GR$) at big scale to explain both $DM$ and $DE$ in the kinematical and dynamical characteristics of stars. Although these both terms can satisfactorily address the problems, they still have significant shortcomings that lead scientists to think about alternate theories of gravity. This has emerged as a flourishing study field in recent years, and we follow a similar pattern in our current work. With the advancement of time, various alterations to the gravitational action of $GR$ have been made. In this manner, the most basic model of altered gravity is $f(R)$ gravity \cite{Sotiriou},\cite{Felice}, where $f(R)$ is a universal function of the Ricci scalar $(R)$. This alteration becomes a part of the game by changing the Einstein-Hilbert action with a random function $f(R)$. Theories with higher-order curvature constants like $f(R)$ and $f(G)$ (where $G$ stands for the Gauss Bonnet invariant), are the modified theories of gravity that have produced some intriguing findings \cite{Nojiri1}-\cite{Nojiri3}. The connection of matter and curvature factors is an intriguing feature of altered theories. Such connection produces a source term that could lead to interesting findings and aid in exploring the puzzles underlying the growth of the cosmos. Encouraged by this assertion, numerous altered theories that encompass well built curvature and matter connection are created like $f(R,T)$ gravity \cite{Harko}, $f(G,T)$ gravity \cite{Haghani} and $f(R,T,R_{ab}T^{ab})$ gravity \cite{Odintsov},\cite{Sharif1}.\\
In \cite{Capozziello}, an elongation of $f(R)$ gravity is given, in which auxiliary degrees of flexibility linked to curvature constants and scalar fields were explored in generalized theories. Surprisingly, such additional levels of flexibility may reformulated as efficient fluids with meanings distinct from the typical matter fluids commonly used as factors in the field equations. Inspired by such findings, the incorporation of matter components into gravitational action was analyzed further in $f(R,T)$ theory, in which the altered action is a random function of $R$ and the trace of energy momentum tensor $T$. Many explorers have investigated cosmic solutions in this framework such as those formed on a isotropic and homogeneous spacetime via a phase space analysis \cite{Shabani}, sturdiness evaluation utilizing energy parameters \cite{Alvarenga,Sharif2}, and thermodynamic features. Its more general implications have been explored in \cite{Sharif3}-\cite{Sharif4}.\\
Furthermore, in \cite{Shabani2}, Shabani and Farhoudi discusses the cosmic outcomes. However the feasibility of various $f(R,T)$ models have already been studied in the cosmic context \cite{Velten}-\cite{Gamonal}, these altered theories can also deal with the compact objects which are categorized in neutron stars $(NS)$, black holes and white dwarfs \cite{sharif}. Compact stars are more massive and have smaller diameters than other ordinary stars. Astrophysicists are provoked to investigate intrinsic characteristics and disparate levels of the formation of compact objects by exploring their nature and specific configurations. Among compact objects, $NS$ have received a lot of attention because of their remarkable structures and characteristics. The most unusual and intriguing objects are $NS$, in which the gravitational pull is countered by the dissipation enforcement of neutrons. The existence of $NS$ was anticipated, right after neutrons were discovered \cite{Baade}. Later on this idea gained significant observational confirmation from the findings of pulsars \cite{Hewish}. Pulsars are observed as the revolving $NS$.\\
The most dense $NS$ in compact stars may further collapse to produce a black hole, while less dense $NS$ may evolve into quark stars $(QS)$. The MIT bag model $(BM)$ equation of state $(EOS)$ is anticipated to be employed to compute internal matter composition of $QS$ \cite{Witten,Cheng}. Several scholars have been inspired in recent years to examine the properties and internal make up of $QS$. In this framework, Rahaman et al. \cite{Rahaman} explored the actual aspects of stars from $6 km$ to the boundary and presented a novel mass function for applicants of strange stars $(SS)$ using the MIT $BM$. With the use of the MIT $BM$ $EOS$, Bhar \cite{Bhar} discussed the attributes of the stars $SAX J 1808.4-3658$, $4U1820-30$ and $PSR J1614-2230$ and discovered their stable structure. By employing the $BM$, Murad \cite{Murad} discussed the impact of charge on anisotropic $SS$ applicants. Arbail and Malheiro \cite{Arbanil} examined the impact of anisotropy in the equilibrium along with the stability of $SS$ by using the numerical results. In this same framework, Deb et al. \cite{Deb1} used the MIT $BM$ $EOS$ to find singularity-free way out of field equations for strange $QS$ and showed that compact stars' anisotropy rises with radial coordinate. It reaches its highest range at the boundary, that appears to be a characteristic of the anisotropic, singularity-free stellar structures.\\
By the use of the MIT $BM$ $EOS$, researchers have been able to better understand the physical characteristics of compact stars and produce some captivating outcomes. With a view to determine the function of matter constituents, Moraes et al. in \cite{Moraes3} addressed the equilibrium configurations of $QS$ using the MIT $BM$. Deb et al. \cite{Deb2} depicted pictorial investigation of the $LMC X-4$ star type by analyzing anisotropic and isotropic source configurations of compact objects. Sharif and Siddiqa \cite{Sharif8} studied the higher curvature terms in frame of $f(R,T)$ gravity for the formation of compact objects by utilizing polytropic and MIT $BM$ $EOS$. Furthermore, Deb et al. \cite{Deb3} assessed the values of the $f(R,T)$ field equations along with anisotropic matter classification for odd $QS$ utilizing the MIT $BM$ $EOS$. Lately, Biswas et al. \cite{Biswas} used the MIT $BM$ $EOS$ for $R+2\phi T$ model to examine the characteristics of some $QS$. Shee and collaborators \cite{Shee} presented a strange star system employing Tolman $V$ type metric potentials utilizing basic form of the MIT $BM$ $EOS$ for the quark matter. Here, we investigate the consequence of the MIT Bag constant $(\mathcal{B})$ on the anisotropic configurations of compact objects for the purpose of exploring the steady structure of $QS$ conforming to the $R+\omega R^{2}+\phi T$ gravity model. In order to explore the stable form of compact things that conform to the $R+\omega R^{2}+\phi T$ gravity model, we examine the influence of $\mathcal{B}$ on the anisotropic arrangement of observed three compact star applicants in this work. \\
The following is our current work scheme. In sec. $2$, we build the formulism for $f(R,T)$ gravity. Further, we talkabout a feasible $f(R,T)$  model and modify the field equations using Tolman V metric potentials relating to anisotropic fluid arrangement in sec. $3$. In sec. $4$, using a strong relationship among inner and outer spacetime for specific terms of the model factors, we determine the solutions of unknown items. The graphical depiction of the physical nature of the examined quark stars is shown in sec. $5$. In the final sec., we gather our findings.
\section{Formulation of $f(R,T)$ GRAVITY}
The action \cite{Harko} presents the $f(R,T)$ gravity, which is expressed as
\begin{equation}\label{1}
  S= \int \Big[\frac{f(R,T)}{2\kappa} + \mathcal{L}_{m}\Big]d^{4}x\sqrt{-g},
\end{equation}
here,
\begin{itemize}
\item $f$, a function, rely on $R$ and $T$,
\item $\mathcal{L}_{m}$ indicate the matter of Lagrangian field,
\item $g$ signifies a metric tensor determinant ($g_{\alpha \beta}$).
\end{itemize}
Corresponding to action given in Eq. (\ref{1}) the field equations are
\begin{equation}\label{2}
f_{R}(R,T)R_{\zeta \eta } - \frac{1}{2}g_{\zeta \eta}f(R,T) - (\nabla_{\zeta}\nabla_{\eta} - g_{\zeta \eta}\Box)f_{R}(R,T) = T_{\zeta \eta}-(T_{\zeta \eta} + \Theta_{\zeta \eta})f_{T}(R,T),
\end{equation}
where $\nabla_{\eta}$ symbolizes covariant derivative, $\Box$ denotes the D'Alembertian symbol, i.e.
$\Box=g^{\zeta \eta}\nabla_{\zeta}\nabla_{\eta}$, $f_{R}(R,T)=\frac{\partial f(R,T)}{\partial R}$, $f_{T}(R,T)=\frac{\partial f(R,T)}{\partial T}$ and $\Theta_{\zeta \eta}$ is assessed by
\begin{equation}\label{3}
\Theta_{\zeta \eta}=g^{u v}\frac{\delta T_{u v}}{\delta g^{\zeta \eta}}=-2T_{\zeta \eta}+g_{\zeta \eta}\mathcal{L}_{m}-2g^{u v} \frac{\partial^{2}\mathcal{L}_{m}}{\partial g^{\zeta \eta}\partial g^{u v}}.
\end{equation}
For interpretation of the intrinsic structure of celestial objects, we observe a static spherically symmetric $(SSS)$ line element provided as
\begin{equation}\label{4}
ds^{2}=e^{\sigma(r)}{dt}^{2}-e^{\gamma(r)}{dr}^{2}-r^{2}({d\theta}^{2}+sin^{2}\theta{d\phi}^{2}),
\end{equation}
Here, ${\sigma (r)}$ and ${\gamma (r)}$ rely on radial coordinate only. \\
In Eq. (\ref{4}),  $T_{\zeta \eta}$ is given as
\begin{equation}\label{5}
 T_{\zeta \eta}=(\rho+p_{r})I_{\zeta}I_{\eta}-p_{t}g_{\zeta \eta}+(p_{r}-p_{t})J_{\zeta}J_{\eta},
\end{equation}
 Here, $I_{\alpha}$ indicates four velocity and $J_{\alpha}$ stands for four vector which satisfy the relations $$J_{\zeta}J^{\zeta}=1,~~~~~I_{\zeta}I^{\zeta}=-1.$$
 Corresponding to fluid distributions, there are many options for matter Lagrangian. In this paper, we take $(\mathcal{L}_{m})=\rho$, here $\rho$ does not rely on the correspondent metric tensor that gives $\frac{\partial^{2}\mathcal{L}_{m}}{\partial g^{\zeta \eta}\partial g^{u v}}=0$ \cite{Harko}. So, now the expression for $\Theta$ is \\ $$\Theta_{\zeta \eta}=-2T_{\zeta \eta}+\rho g_{\zeta \eta}.$$
 After inserting $\Theta_{\zeta \eta}$ into Eq. (\ref{2}), we obtain
 \begin{equation}\label{6}
 G_{\zeta \eta}=\frac{1}{f_{R}}\Big[T_{\zeta \eta}(1+f_{T})+\frac{1}{2}(f-R f_{R})g_{\zeta \eta}-\rho g_{\zeta \eta}f_{T}-(g_{\zeta \eta}\Box -\nabla_{\zeta}\nabla_{\eta})f_{R}\Big].
 \end{equation}
For $f(R,T)$ gravity in Eq. (\ref{6}), below are the field equations:
\begin{equation}\label{7}
\rho= e^{-\gamma}\Big[\Big(\frac{\sigma'^{2}}{4}+\frac{\sigma''}{2}+\frac{\sigma'}{r}-\frac{\sigma'\gamma'}{4}\Big)f_{R}+\Big(\frac{\gamma'}{2} - \frac{2}{r}\Big)f'_{R} - f''_{R}\Big]-\frac{f}{2},~~~~~~~~~~~~~~~~~~
\end{equation}
\begin{equation}\label{8}
p_{r} = \frac{e^{-\gamma}}{1+f_{T}}\Big[\Big(\frac{\gamma'}{r}-\frac{\sigma'^{2}}{4}-\frac{\sigma''}{2}+\frac{\sigma' \gamma'}{4}\Big)f_{R}+\Big(\frac{\sigma'}{2}+\frac{2}{r}\Big)f'_{R}\Big]-\rho f_{T}+\frac{f}{2},~~~~~~~~~~
\end{equation}
\begin{equation}\label{9}
p_{t}=\frac{e^{-\gamma}}{1+f_{T}}\Big[\Big(\frac{\gamma'}{2r}-\frac{\sigma'}{2r}-\frac{1}{r^{2}}+\frac{e^{\gamma}}{r^{2}}\Big)f_{R}+\Big(\frac{\sigma'-\gamma'}{2}+\frac{1}{r}\Big)f'_{R} +f''_{R}\Big]-\rho f_{T}+\frac{f}{2},
\end{equation}
here prime exhibits derivative respecting radial coordinate.
\section{Viable $f(R,T)$ Gravity Model}
It has been demonstrated that a number of star systems are now existent in a nonlinear system at the current cosmic stage. We have to examine their linear behaviour to get a whole view of their structure transformation. In order to talk about how the relativistic structures are affected by the link of curvature and matter  constituents in $f(R,T)$ gravity, We take a divisible functional expression proposed by
\begin{equation}\label{10}
f(R,T)=f_{1}(R)+h(T).
\end{equation}
This type of separable models may give an appropriate linear extension of $f(R)$ gravity. This model can be used to create a variety of viable $f(R,T)$ gravity models on selecting various $f_{1}(R)$ configurations together with linear combinations of $h(T)$. Here, we consider $h(T)=\phi T$, where $T=\rho-p_{r}-2p_{t}$ and $\phi$ is a positive constant. In this aspect, the field Eqs. (\ref{7})-(\ref{9}) turn out to be\\
\begin{equation}\label{11}
\Big(1+\frac{\phi}{2}\Big)\rho-\frac{\phi}{2}p_{r}-\phi p_{t}= e^{-\gamma}\Big[\Big(\frac{\sigma'^{2}}{4}+\frac{\sigma''}{2}+\frac{\sigma'}{r}-\frac{\sigma'\gamma'}{4}\Big)f_{1R}+\Big(\frac{\gamma'}{2} - \frac{2}{r}\Big)f'_{1R} - f''_{1R}-\frac{e^{\gamma}f_{1}}{2}\Big],~~~~~~~
\end{equation}
\begin{equation}\label{12}
\frac{\phi}{2}\rho+\Big(1+\frac{3\phi}{2}\Big)p_{r}+\phi p_{t}
=e^{-\gamma}\Big[\Big(\frac{\gamma'}{r}-\frac{\sigma'^{2}}{4}-\frac{\sigma''}{2}+\frac{\sigma' \gamma'}{4}\Big)f_{1R}+\Big(\frac{\sigma'}{2}+\frac{2}{r}\Big)f'_{1R}+\frac{e^\gamma f_{1}}{2}\Big],~~~~~~~
\end{equation}
\begin{equation}\label{13}
\frac{\phi}{2}\rho+\frac{\phi}{2}p_{r}+(1+2\phi)p_{t}=e^{-\gamma}\Big[\Big(\frac{\gamma'}{2r}-\frac{\sigma'}{2r}-\frac{1}{r^{2}}+\frac{e^{\gamma}}{r^{2}}\Big)f_{1R}+\Big(\frac{\sigma'-\gamma'}{2}+\frac{1}{r}\Big)f'_{1R} +f''_{1R}+e^\gamma\frac{f_{1}}{2}\Big].
\end{equation}\\
In the relativistic model of $QSs$, we propose that the MIT $BM$ $EOS$ influences fluid dispersion in the stars' cores \cite{Witten,Cheng}. The quark pressure is determined as below
\begin{equation}\label{14}
p_{r}=\sum_{j=U,D,S} p^{j}-\mathcal{B}.
\end{equation}
Here, $\mathcal{B}$ is bag constant and $p^{j}$ denotes the corresponding pressure of down $(D)$, up $(U)$ and strange $(S)$ quark indications. The independent quark pressure and energy density of the corresponding quark flavors are correlated as $p^{j}=\frac{1}{3}\rho^{i}$. So, energy density is given as
\begin{equation}\label{15}
\rho=\sum_{j=U,D,S} p^{j}+\mathcal{B}.
\end{equation}
By Eqs. (\ref{14}) and (\ref{15}), MIT $BM$ $EOS$ for quark stuff is formulated as
\begin{equation}\label{16}
p_{r}=\frac{1}{3}(\rho-4\mathcal{B}).~~
\end{equation}
It has been noted that various researchers have effectively employed the simplest type of this $EOS$ to study the characteristics of $QS$ applicants.  We employ the Tolman V metric potentials, to examine physical features of compact relativistic stars, given below \cite{Hansraj}.
\begin{equation}\label{17}
\sigma(r)=ln[Yr^{2b}],
\end{equation}
\begin{equation}\label{18}
\gamma(r)=ln\Big[\frac{1+2b-b^{2}}{1-(1+2b-b^{2})(\frac{r}{F})^{B}}\Big],
\end{equation}
here, $B=\frac{2(1+2b-b^{2})}{b+1}$, $b$, $Y$ and (say, $(\frac{1}{F})^{B}=A$) are constants.
As a result, the Tolman V space-time clearly appears as
\begin{equation}\label{19}
ds^{2}=\Big[Yr^{2b}\Big]dt^{2}-\Big[\frac{1+2b-b^{2}}{1-(1+2b-b^{2})(r)^{B}A}\Big]dr^{2}-r^{2}d\Omega^{2}.
\end{equation}\\
By determining the metric potentials, we ensure that the potentials in the compact object are well behaved, stable and non singular.

Now, with in the context of Eq. (\ref{17}) and Eq. (\ref{18}), Eqs. (\ref{11}) - (\ref{13}) along with Eq. (\ref{16}) become\\
\begin{equation}\label{20}
\rho=\frac{3(1-A(w_6)r^{B})}{4(w_6)(1+\phi)}\Big[(w_1)f_{1R}+(w_2)f'_{1R}-f''_{1R}\Big]+\mathcal{B},~~~~~~~~~~~~~~~~~~~~~~~~~~~~~~~~~~~~~~~~~~~~~~~~~~~~~~~~~~
\end{equation}
\begin{equation}\label{21}
p_{r}=\frac{1-A(w_6)r^{B}}{4(w_6)(1+\phi)}\Big[(w_1)f_{1R}+(w_2)f'_{1R}-f''_{1R}\Big]-\mathcal{B},~~~~~~~~~~~~~~~~~~~~~~~~~~~~~~~~~~~~~~~~~~~~~~~~~~~~~~~~~~~~
\end{equation}
\begin{equation}\label{22}
p_{t}=1-A(w_6)r^{B}\Big[-\frac{\phi((w_1)f_{1R}+(w_2)f'_{1R}-f''_{1R})}{2(w_5)(1+\phi)}+\frac{(w_3)f_{1}+(w_4)f_{1R}+(w_5)f'_{1R}+f''_{1R}}{(w_6)(1+2\phi)}\Big],~~~~~~~~~~~~
\end{equation}
where, \\
$w_1=\frac{2b}{r^{2}}+\frac{A(w_6)Br^{-2+B}}{1-A(w_6)r^{B}}$, \\ \\
$w_2=\frac{b}{r}+\frac{A(w_6)Br^{-1+B}}{2(1-A(w_6)r^{B})}$,\\ \\
$w_3=\frac{(1-A(w_6)r^{B}}{2(w_6)}$,\\ \\
$w_4=-\frac{1}{r^{2}}-\frac{b}{r^{2}}+\frac{(w_6)}{r^{2}(1-A(w_6)r^{B})}+\frac{A(w_6)Br^{-2+B}}{2(1-A(w_6)r^{B}}$,\\ \\
$w_5=\frac{1}{r}+\frac{b}{r}-\frac{A(w_6)Br^{-1+B}}{2(1-A(w_6)r^{B}}$,\\ \\
$w_6=1+2b-b^{2}$.\\ \\
The expression $p_{r}(r=R)=0$ shows the dispersion of $p_{r}$ at surface boundary in the stellar structure and is used to find the evaluation of $\mathcal{B}$ which is
\begin{equation}\label{23}
\mathcal{B}=\frac{1-A(w_6)r^{B}}{4(w_6)(1+\phi)}\Big[(w_1)f_{1R}+(w_2)f'_{1R}-f''_{1R}\Big].
\end{equation}\\
Now, in the environment of specific visible survey, we explore the effect of our given model on the systemic development and steadiness of compact stars. Various $f(R,T)$  models possibly used as mathematical devices to look at a variety of hidden aspects of gravitational changing at enormous levels. Comparing with $f(R)$ gravity, the changing of $f(R,T)$ gravity contain an extension of $T$ which describes a very extended type of $GR$. The selection of astrophysical functional forms is governed by the requirements for cosmological consistency set out by the solar system tests. One can create numerous $f(R,T)$ models regulated by the possibility of $f_1(R)$ provided in Eq. (\ref{10}). \\
The extension of quadratic $R$, first proposed by Starobinsky \cite{Starobinsky}, is taken into consideration in this paper. The $f(R,T)$ model in Eq. (\ref{10}), after this modification, set off
\begin{equation}\label{24}
f(R,T)= R+\omega R^2+\phi T,
\end{equation}
where $\omega$ is a freely chosen parameter. The above model effectively explains the present expanding in the universe's expansion, and it can be examined as a different possibility for $DE$. The $GR$ field equations can be generated by adding $\omega=0=\phi$   to this model. This model is frequently used in literature to analyze both gravitational collapse and stellar evolution. This model was developed by Moraes et al. \cite{Moraes4}, who discovered that its functional form effectively portrays the stellar dilemma of a radiation dominated cosmos. For this functional form, Sharif and Siddiqa \cite{Sharif8} studied spherically symmetric astronomical shape applying both the polytropic and the MIT $BM$ $EOS$.
\section{Matching Conditions}
The entire framework of the self gravitating compact stellar object is shown in this section by smoothly matching our internal metric with the external metric at the pressure free boundary ($r=R$). For this objective, we used the Schwarzschild metric, that accurately expresses the outer area of compact bodies defined as
\begin{equation}\label{25}
ds^{2}=\Big[1-\frac{2M}{R}\Big]dt^{2}-\Big[1-\frac{2M}{R}\Big]^{-1}dr^{2}-r^{2}d\Omega^{2},
\end{equation}
where the entire mass is denoted by $M$. The following limitations are produced by the continuum of the metric variables $g_{tt,R}$, $g_{tt}$ and $g_{rr}$ at the edges connecting the interior and exterior spacetimes of compact objects.
\begin{equation}\label{26}
2YbR^{-1+2b}=\frac{2M}{R^{2}},~~~~~~~~~~~~~~~~~~~~~~~~~~~~
\end{equation}
\begin{equation}\label{27}
YR^{2b}=1-\frac{2M}{R},~~~~~~~~~~~~~~~~~~~~~~~~~~~~~~
\end{equation}
\begin{equation}\label{28}
\frac{1+2b-b^{2}}{1-(1+2b-b^{2})(R)^{B}A}=\Big(1-\frac{2M}{R}\Big)^{-1}.
\end{equation}\\
The formulations of undetermined constants $Y$, $b$ and $A$ in the mass and radius combination possibly found by resolving Eqs. (\ref{26})-(\ref{28}) which gives
\begin{equation}\label{29}
Y=\frac{M}{R^{1+2b}b}, ~~~~~~~~~~~~~~~~~~~~~~~~~~~~~~~~~~
\end{equation}
\begin{equation}\label{30}
b=\frac{M}{R-2M},~~~~~~~~~~~~~~~~~~~~~~~~~~~~~~~~~
\end{equation}
\begin{equation}\label{31}
A=\frac{M(MR-2M^{2})}{R^{1+B}(R^{2}-2MR-M^{2})}.~~~~~~~~~~~~~
\end{equation}\\
For $\omega=0$, by using Eq. (\ref{24}) and by putting values of Eqs. (\ref{29})-(\ref{31}), Eq. (\ref{23}) simplifies to $\mathcal{B}$ for the most basic $R+2\phi T$ gravity model \cite{Biswas}. Here, by using the radii and masses of the $SS$ objects listed in \cite{Bombaci1}-\cite{X. D. Li}, the values of $Y$, $b$ and $A$ can be measured. These celestial objects are in accordance with the limit proposed by Buchdahl, which states that, the rates of $\frac{2M}{R}$ must be lesser than $\frac{8}{9}$ \cite{Buchdahl}. We choose $\omega=-15, -55, -95$ and $\phi=-1.1$ to assess the values of unknown parameters for our suggested stars. These model parameter values have been used to examine the formation of stellar structures successfully. We have shown the values of $Y$, $b$ and $A$ according to the radius and mass values of the observed stars in  TABLE $\ref{Table 1}$. $\mathcal{B}$ can take on a wide range of values for a density reliant $BM$, according to experimental findings from $CERN-SPS$ and $RHIC$ \cite{Burgio}.
 
\begin{table}[h!]
\caption{Estimated values of the unknown parameters $Y$, $b$ and $A$ for $RXJ 1856-37$ \cite{Arfa}, $Her X-1$ \cite{Arfa},  $LMC X-4$ \cite{Zoya}.}
\centering
\begin{tabular}{|p{4.2cm}| p{2.3cm}| p{1.4cm}| p{2.0cm}| p{2.0cm}| p{1.6cm}| p{2.6cm}|}
\hline
\hline
\begin{center}
  Objects
\end{center} & ~~~\begin{center}
    $M(M_\odot)$
    \end{center} & ~~~ \begin{center}
    $R(km)$
    \end{center} & ~~~ \begin{center}
    $n(km)$
    \end{center} & ~~~ \begin{center}
    $B(km)$
    \end{center} & ~~~ \begin{center}
    $W(km)$
    \end{center} & ~~~ \begin{center}
    $A(km)$
    \end{center}\\
    \hline
~~~~$RXJ 1856-37~~(s1)$ &~~0.9041$(M_\odot)$	&~~~~6	&~~~0.404094 &~~~0.129976  &~~~~2.343  &~~~0.000824849\\\hline

~~~~~ $Her X-1~~(s2)$ &~~~0.88$(M_\odot)$	&~~~7.7	&~~~0.256426	&~~~0.232041 &~~2.30351 &~~~0.000272638\\\hline

~~~~~ $LMC X-4~~(s3)$ &~~~1.04$(M_\odot)$	&~~~8.301	&~~~0.295705	&~~~0.179736 &~~2.32147 &~~~0.000268513\\
\hline
\end{tabular}
\label{Table 1}
\end{table}
 
The physical conditions that are crucial for understanding the inner region of stellar models are described in the following section.
\section{Physical Acceptability Conditions of Strange Stars}
This section examines numerous physical characteristics of noted anisotropic dense stellar possibilities. We get the values of the matter parameters by putting the measured data from Table TABLE $\ref{Table 1}$ with the suggested model (\ref{24}). We investigate to see how the energy density, transverse and radial pressures, energy constraints, anisotropic factor, compactness, redshift factor, and stability behave graphically for specific values of the model's components. In the framework of theoretical and astrophysical scenarios, this graphical analysis could reveal certain enigmatic realities.

\subsection{Formation of Matter Variables}
Compact objects have the largest densities, thus it follows that they will have the greatest impact on the components of the star system's energy density and pressure. For our suggested model, Figs. $(\ref{1})$-$(\ref{3})$ show how these features inside the proposed compact objects vary  respecting  radial coordinate. These figures obviously show how the energy density and pressure constituents exhibit peak  rates near  the core of anisotropic compact objects, resulting in the existence of really compact centers. Additionally, it is discovered that the radial pressure diminishes at the boundary, the density and pressure constituents are functions of the radial coordinate which are decreasing, and they have definite values in  the stars that match the estimated rates of $\mathcal{B}$.

\begin{figure}[h!]
\begin{tabular}{cccc}
\epsfig{file=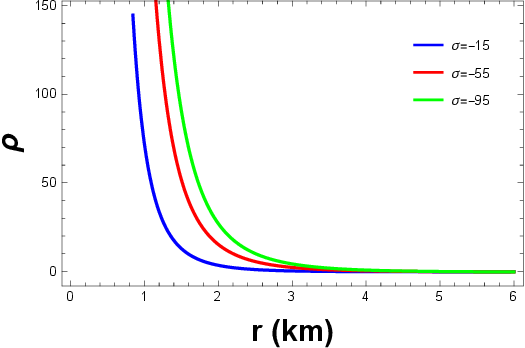, width=.30\linewidth} &
\epsfig{file=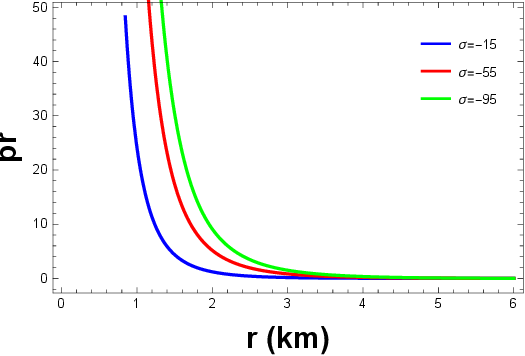, width=.30\linewidth} &
\epsfig{file=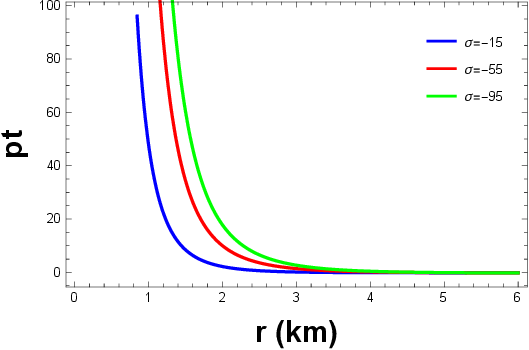, width=.30\linewidth} &
\end{tabular}
\caption{{Variation of $\rho$, $p_r$ and $p_t$} for $s1$.}
\label{Fig.1}
\end{figure}
 
\begin{figure}[h!]
\begin{tabular}{cccc}
\epsfig{file=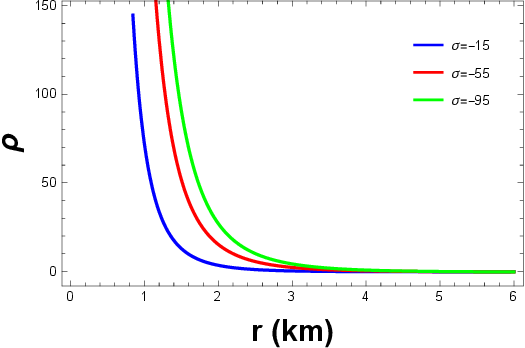, width=.30\linewidth} &
\epsfig{file=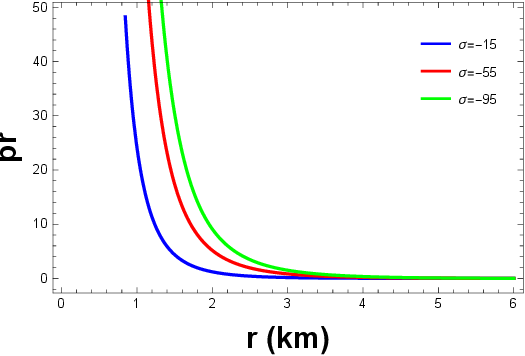, width=.30\linewidth} &
\epsfig{file=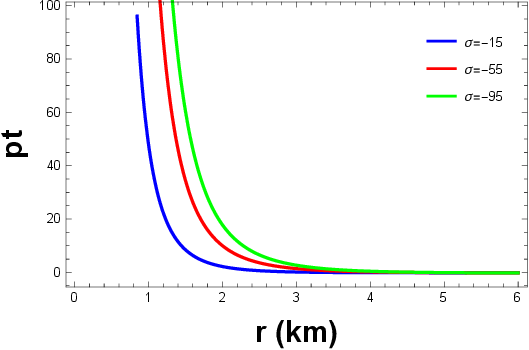, width=.30\linewidth} &
\end{tabular}
\caption{{Variation of $\rho$, $p_r$ and $p_t$} for $s2$.}
\label{Fig.2}
\end{figure}

\begin{figure}[h!]
\begin{tabular}{cccc}
\epsfig{file=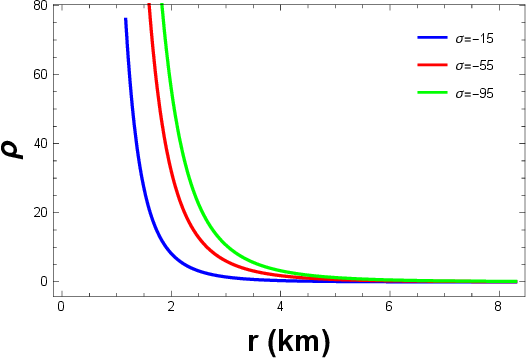, width=.30\linewidth} &
\epsfig{file=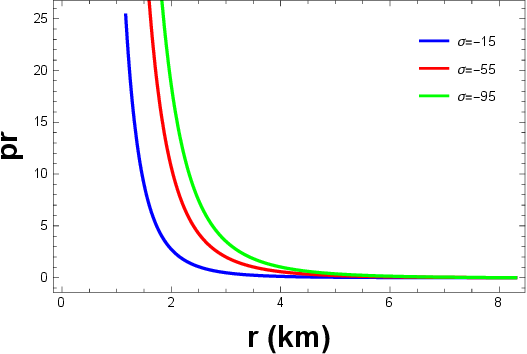, width=.30\linewidth} &
\epsfig{file=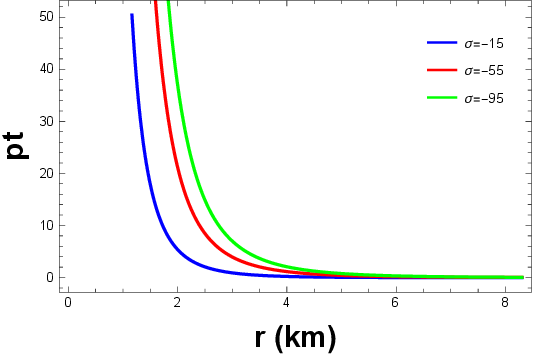, width=.30\linewidth} &
\end{tabular}
\caption{Variation of $\rho$, $p_r$ and $p_t$ for $s3$.}
\label{Fig.3}
\end{figure}

The results of this visible nature offer a very firm outline of the stars under consideration, ensuring that dense stars exist. \\
Figs. $(\ref{4})$-$(\ref{6})$ illustrate how the rate of change in energy density ($\frac{d\rho}{dr}$), radial pressure ($\frac{dp_{r}}{dr}$), and tangential pressure ($\frac{dp_{t}}{dr}$) of $s1$, $s2$, $s3$ is zero at the core and gets negative as it approaches the bound.\\
\begin{figure}[h!]
\begin{tabular}{cccc}
\epsfig{file=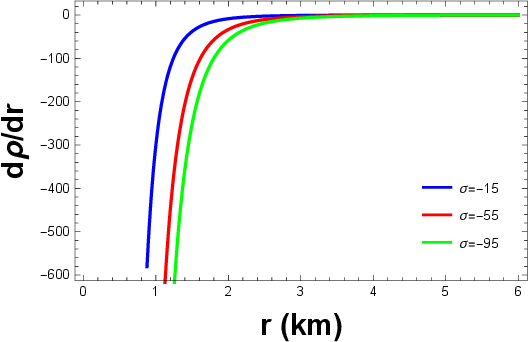, width=.30\linewidth} &
\epsfig{file=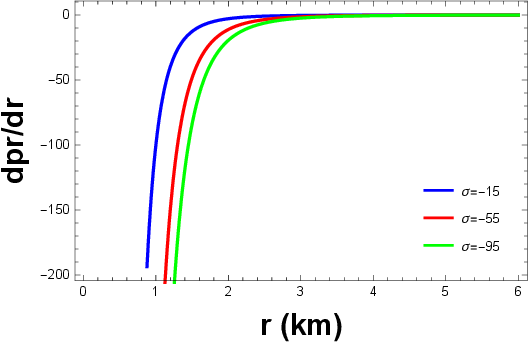, width=.30\linewidth} &
\epsfig{file=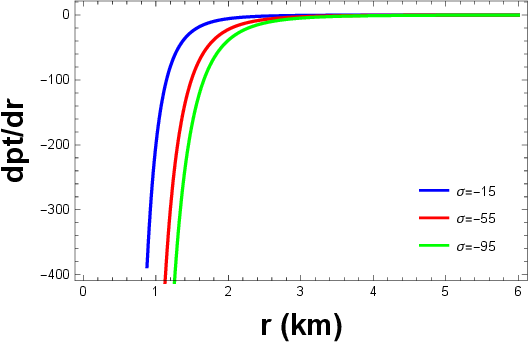, width=.30\linewidth} &
\end{tabular}
\caption{Graphs of $\frac{d\rho}{dr}$, $\frac{dp_{r}}{dr}$ and $\frac{dp_{t}}{dr}$ for $s1$.}
\label{Fig.4}
\end{figure}

\begin{figure}[h!]
\begin{tabular}{cccc}
\epsfig{file=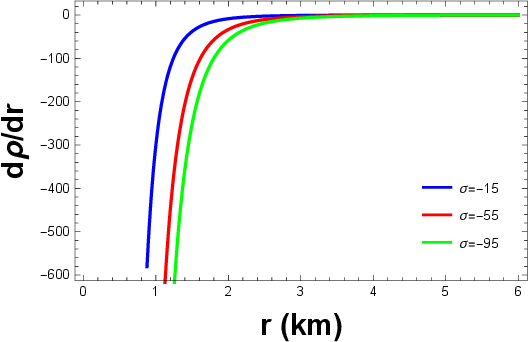, width=.30\linewidth} &
\epsfig{file=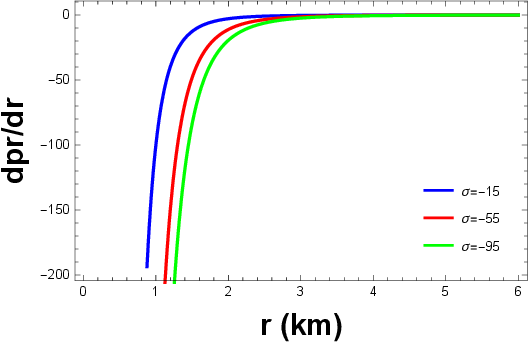, width=.30\linewidth} &
\epsfig{file=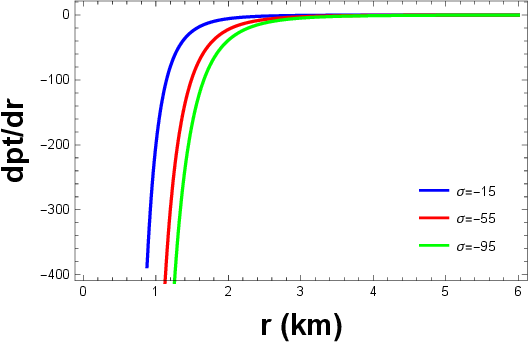, width=.30\linewidth} &
\end{tabular}
\caption{Graphs of $\frac{d\rho}{dr}$, $\frac{dp_{r}}{dr}$ and $\frac{dp_{t}}{dr}$ for $s2$.}
\label{Fig.5}
\end{figure}

\begin{figure}[h!]
\begin{tabular}{cccc}
\epsfig{file=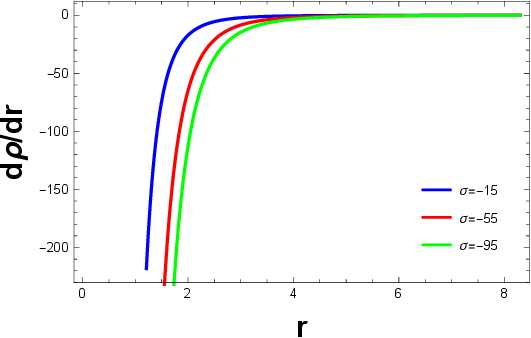, width=.30\linewidth} &
\epsfig{file=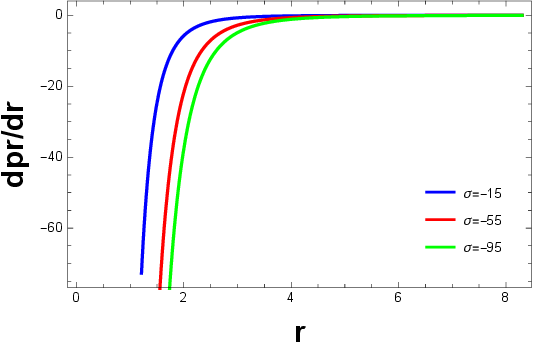, width=.30\linewidth} &
\epsfig{file=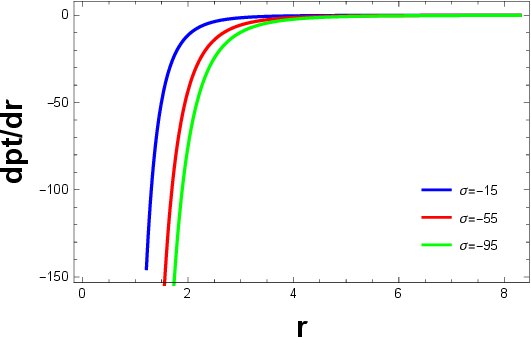, width=.30\linewidth} &
\end{tabular}
\caption{Graphs of $\frac{d\rho}{dr}$, $\frac{dp_{r}}{dr}$ and $\frac{dp_{t}}{dr}$ for $s3$.}
\label{Fig.6}
\end{figure}
 
\subsection{Anisotropy Parameter}
In order to obtain a better knowledge of interior structure of relativistic compact stars, we use an anisotropic factor, which is formulated as
\begin{equation}\label{32}
\Delta=p_{t}-p_{r}.
\end{equation}
Along the aid of observed information of the specified anisotropic compact objects, shown in TABLE $\ref{Table 1}$, we graphically evaluate the behavior of anisotropy. If $p_{t}>p_{r}$, the result is $\Delta>0$, which depicts the outward pointed anisotropic pressure, while $p_{t}<p_{r}$ results in $\Delta<0$, which indicates the inward pointed anisotropic pressure.

\begin{figure}[h!]
\begin{tabular}{cccc}
\epsfig{file=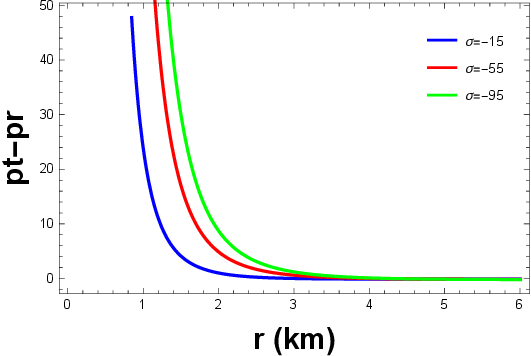, width=.30\linewidth} &
\epsfig{file=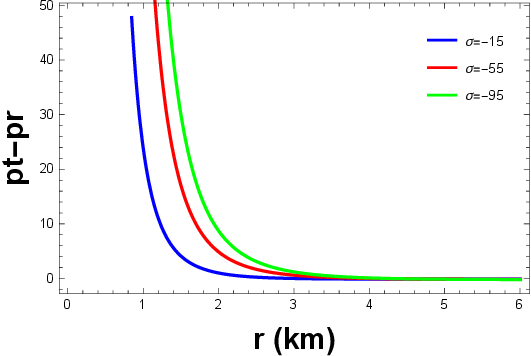, width=.30\linewidth} &
\epsfig{file=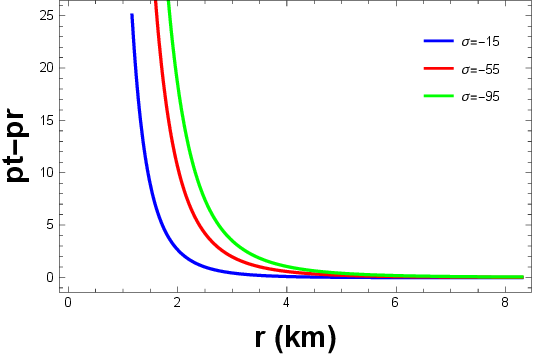, width=.30\linewidth} &
\end{tabular}
\caption{Pictorial Representation of anisotropy for $s1$, $s2$ and $s3$.}
\label{Fig.7}
\end{figure}
 
Fig. $(\ref{7})$ depict the outcome of the anisotropic assessment \cite{Zubair} for compact star candidates in correspondence with the usable functional form of the $f(R,T)$ gravity. The variation of $\Delta$ is observed to be positive, indicating the existence of a repulsive force that permits for the emergence of more dense dispersion in the interior configuration of star models.
\subsection{Energy Conditions}
There are physical characteristics which are called energy conditions that can be used to explore the existence of an actual matter dispersion. These circumstances are essential for observing whether the stuff in the interior of the star model is regular or unusual.
There are four types within these energy states: weak, strong, null and dominant.
%Such states in the existence of anisotropic fluid for curvature-matter coupled gravity are given as \cite{Gasperini}
\begin{itemize}
\item $WEC$: ~~~~~$\rho-C \geq 0$,~~~~~$\rho+p_{r}-C \geq 0$,~~~~~~~~~~~~~$\rho+p_{t}-C \geq 0$,
\item $SEC$: $\rho+p_{r}-C \geq 0$,~~~~ $\rho+p_{t}-C \geq 0$,~~~~
$\rho+p_{r}+2p_{t}-C \geq 0$,
\item $NEC$: $\rho+p_{r}-C \geq 0$,~~~~ $\rho+p_{t}-C \geq 0$,
\item $DEC$: $\rho-p_{r}-C \geq 0$,~~~~  $\rho-p_{t}-C \geq 0$,
\end{itemize}
where $C=\nabla_{u}(V^{v}\nabla_{v}V^{u})$ emerges because of the heavy fragments' non geodesic motion and is described as
$$C=e^{-\gamma}\Big(\frac{\sigma'}{r}+\frac{\sigma''}{2}+\frac{\sigma'^{2}}{4}-\frac{\gamma'\sigma'}{4}\Big).$$
The energy conditions in Figs. $(\ref{8})$-$(\ref{10})$ are plotted, applying Tolman V metric potentials in the representation of $C$. For the researched compact stars, it has been observed that all energy requirements have been fulfilled, confirming the survival of usual matter near to the interior of quark star applicants. As a result, the anisotropy and $\mathcal{B}$ describe the actual origination of the gravitational forces of celestial bodies.
\begin{figure}[h!]
\begin{tabular}{cccc}
\epsfig{file=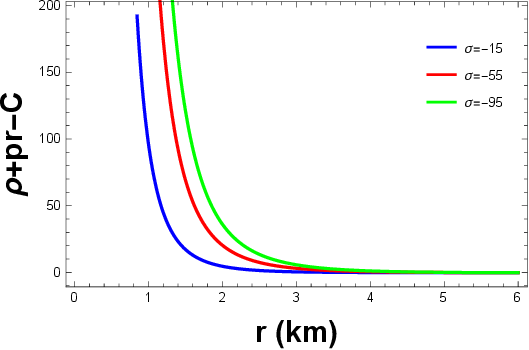, width=.30\linewidth} &
\epsfig{file=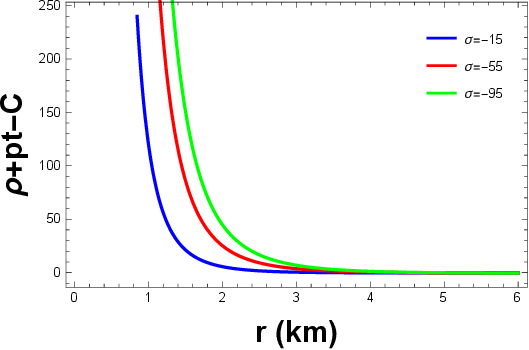, width=.30\linewidth} &
\epsfig{file=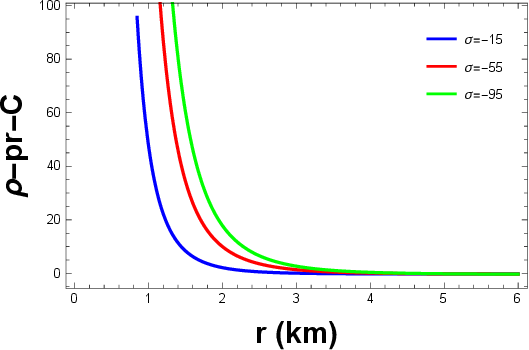, width=.30\linewidth} &\\\\\\\\
\end{tabular}
\begin{tabular}{cccc}
\epsfig{file=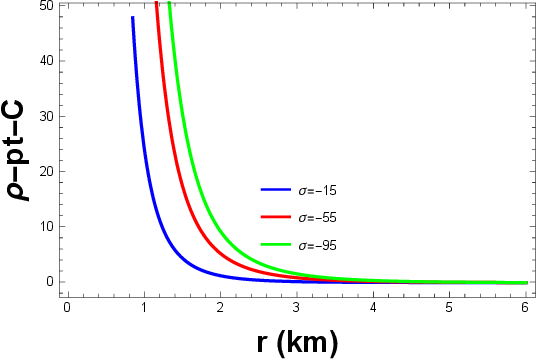, width=.30\linewidth}
\epsfig{file=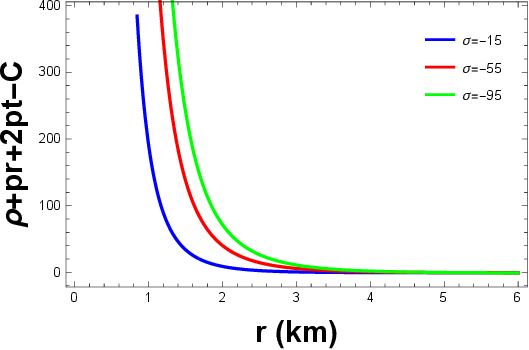, width=.30\linewidth} &
\epsfig{file=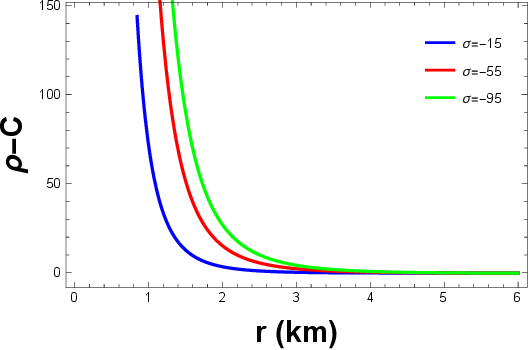, width=.30\linewidth} &\\\\\\\\
\end{tabular}
\caption{{Graphs of energy conditions for $s1$  against $r$}.}
\label{Fig.8}
\end{figure}

\begin{figure}[h!]
\begin{tabular}{cccc}
\epsfig{file=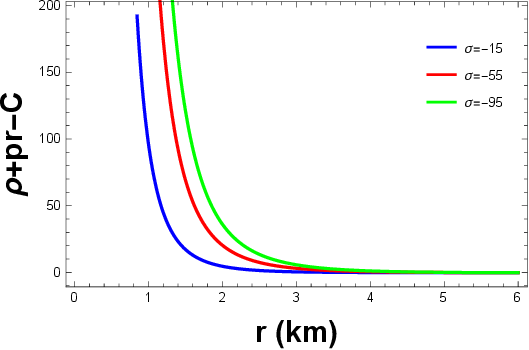, width=.30\linewidth} &
\epsfig{file=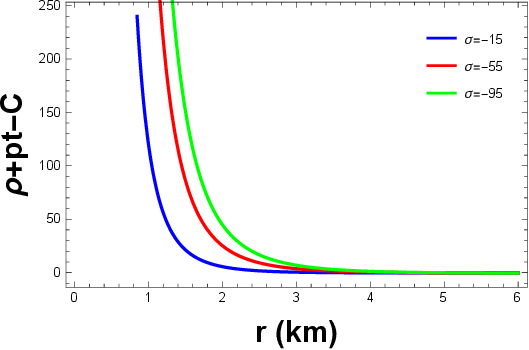, width=.30\linewidth} &
\epsfig{file=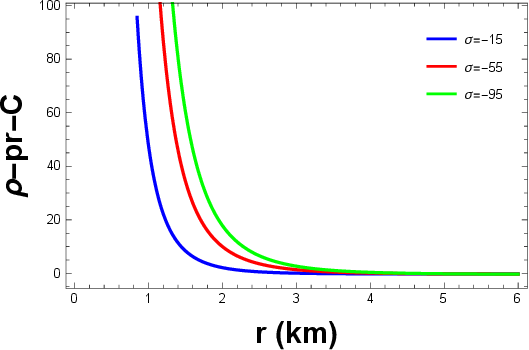, width=.30\linewidth} &\\\\\\\\
\end{tabular}
\centering
\begin{tabular}{cccc}
\epsfig{file=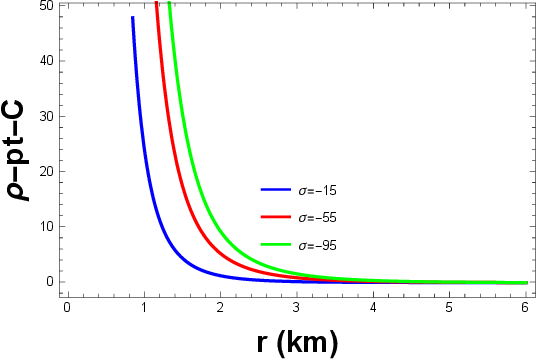, width=.30\linewidth} &
\epsfig{file=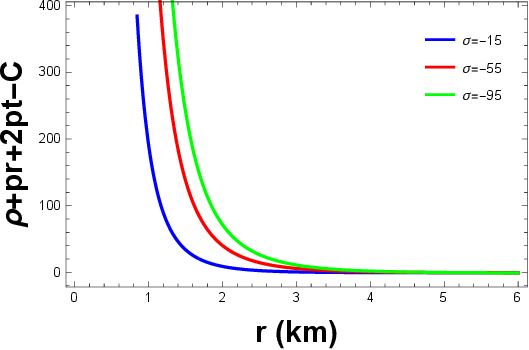, width=.30\linewidth} &
\epsfig{file=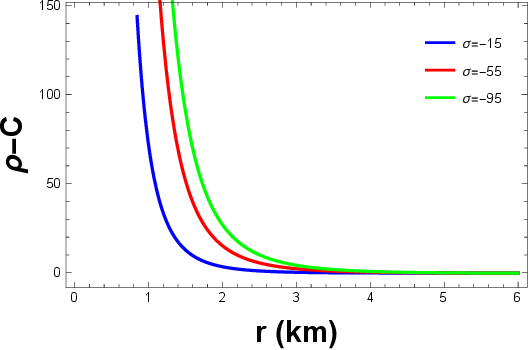, width=.30\linewidth} &\\\\\\\\
\end{tabular}
\caption{{Graphs of energy conditions for $s2$  against $r$}.}
\label{Fig.9}
\end{figure}

\begin{figure}[h!]
\begin{tabular}{cccc}
\epsfig{file=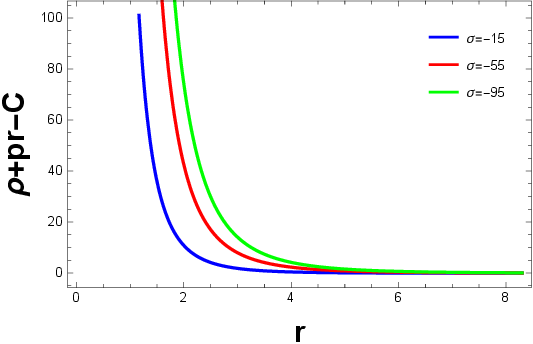, width=.30\linewidth} &
\epsfig{file=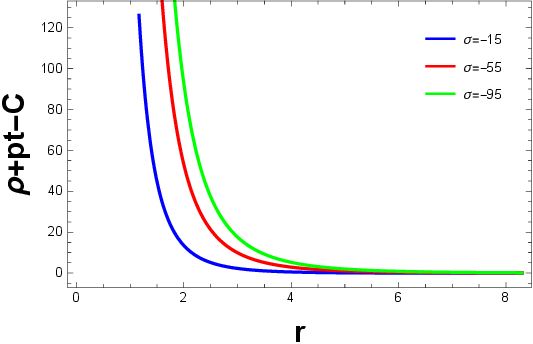, width=.30\linewidth} &
\epsfig{file=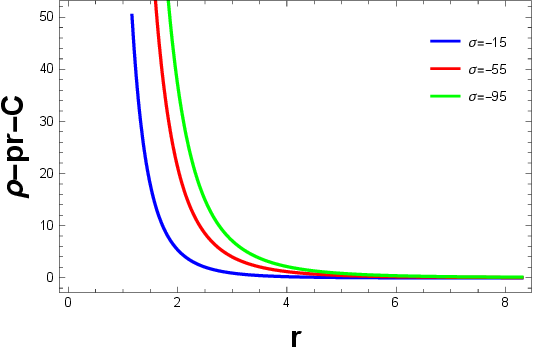, width=.30\linewidth} &\\\\\\\\
\end{tabular}
\centering
\begin{tabular}{cccc}
\epsfig{file=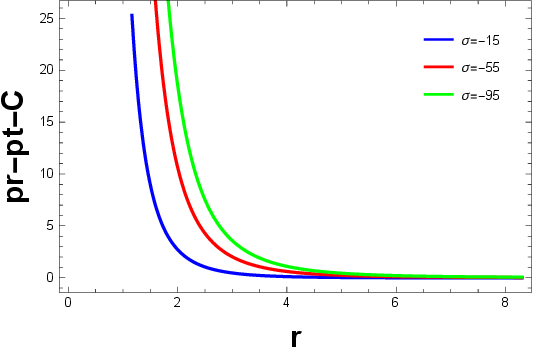, width=.30\linewidth} &
\epsfig{file=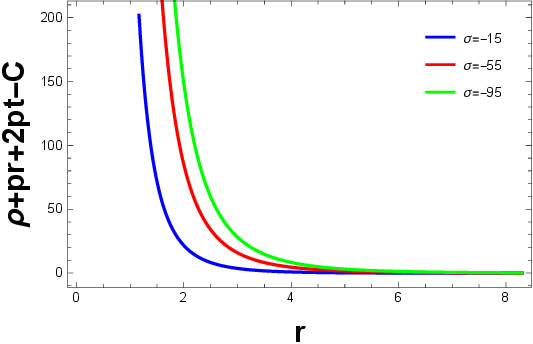, width=.30\linewidth} &
\epsfig{file=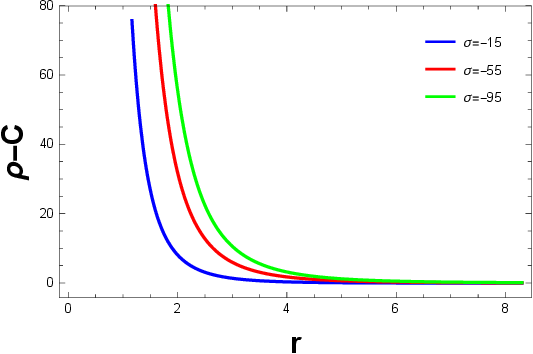, width=.30\linewidth} &\\\\\\\\
\end{tabular}
\caption{{Graphs of energy conditions for $s3$  against $r$}.}
\label{Fig.10}
\end{figure}
 
\subsection{Mass-Radius association, Compactness Component and Surface Red-shift}
To discuss the stability of compact star, mass function, compactness factor and red-shift function are key features.
All these attributes are connected with each other. Buchdahl \cite{Buchdahl} discovered a limitation on the mass radius ratio, i.e. $\frac{2M}{R}<\frac{8}{9}$, for $SSS$ celestial models influenced by anisotropic
fluid. The mass function for the compact star is given as
\begin{equation}\label{44}
M=\frac{R}{2}(1-e^{-\gamma}),
\end{equation}

From the Figs. $(\ref{11})$-$(\ref{13})$, it can be seen that the mass of the sphere has a direct relationship with its radius,
indicating that mass is conventional inside the star i.e $M\rightarrow 0$, $R\rightarrow 0$. The compactness component, known as mass-radius ratio,
given as

\begin{equation}\label{45}
u=\frac{M}{R}.
\end{equation}
Understanding the solid visible contacts among fragments within  the star and its $EOS$ depends on surface red-shift.
The surface red-shift $(z_{s})$, is defined as
\begin{equation}\label{46}
z_{s}=\frac{1}{\sqrt{1-2u}}-1.
\end{equation}
\begin{figure}[h!]
\begin{tabular}{cccc}
\epsfig{file=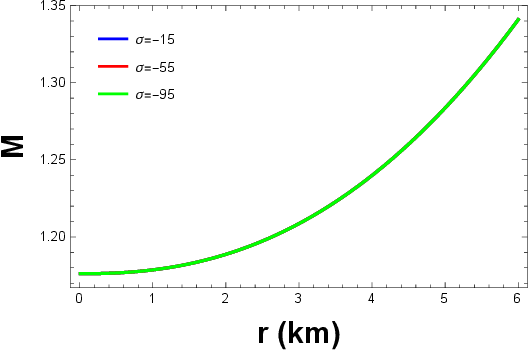, width=.30\linewidth} &
\epsfig{file=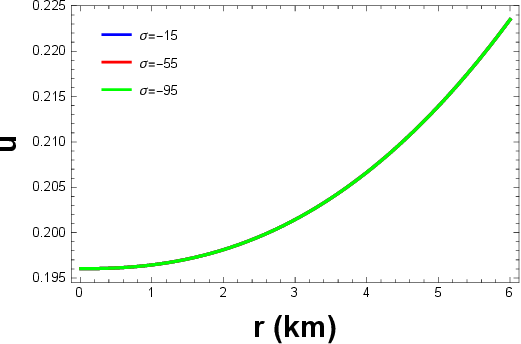, width=.30\linewidth} &
\epsfig{file=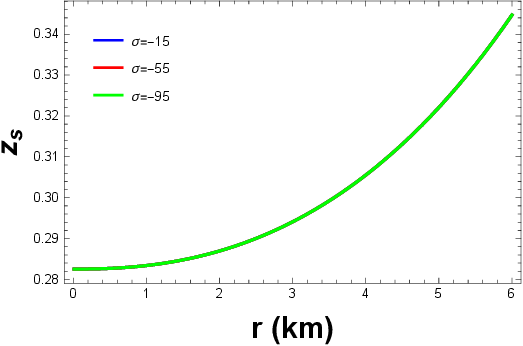, width=.30\linewidth} &
\end{tabular}
\caption{{Nature of mass, compactness component and surface red-shift for $s1$}.}
\label{Fig.11}
\end{figure}

\begin{figure}[h!]
\begin{tabular}{cccc}
\epsfig{file=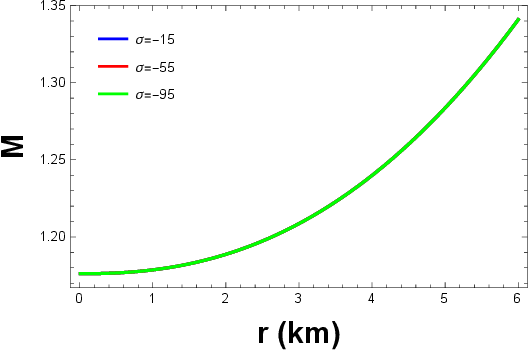, width=.30\linewidth} &
\epsfig{file=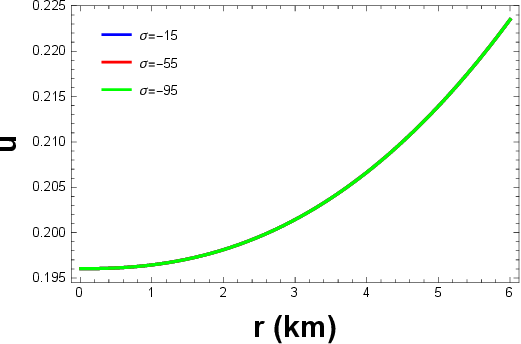, width=.30\linewidth} &
\epsfig{file=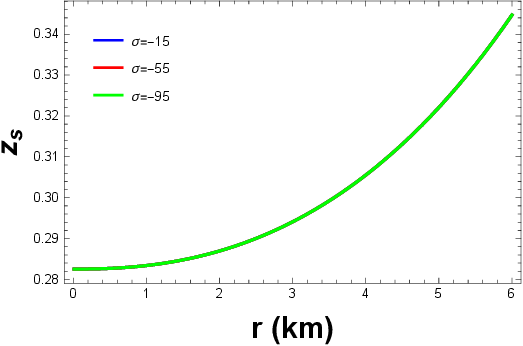, width=.30\linewidth} &
\end{tabular}
\caption{{Nature of mass, compactness component and surface red-shift for $s2$}.}
\label{Fig.12}
\end{figure}

\begin{figure}[h!]
\begin{tabular}{cccc}
\epsfig{file=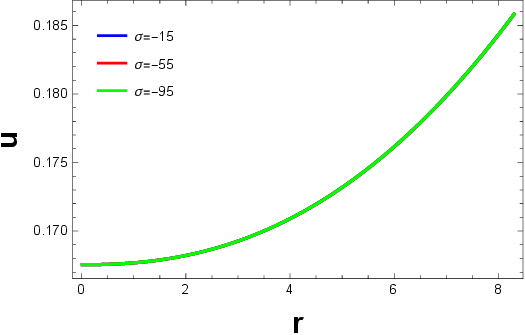, width=.30\linewidth} &
\epsfig{file=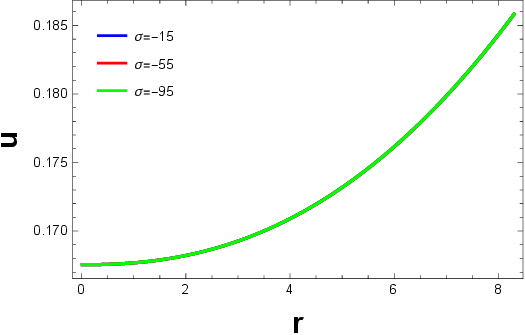, width=.30\linewidth} &
\epsfig{file=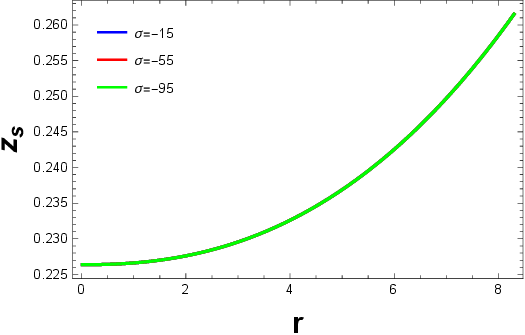, width=.30\linewidth} &
\end{tabular}
\caption{{Nature of mass, compactness component and surface red-shift for $s3$}.}
\label{Fig.13}
\end{figure}
 
Surface redshift is a characteristic of the strong intermolecular interactions between the particles in the core of the star. The compactness of the star objects will rise when the forces of attraction rise. The pictorial depiction in  Figs. $(\ref{11})$-$(\ref{13})$ illustrates how the compactness factor and surface red-shift have evolved over time. Both evolutions rise as they proceed towards the border and diminish as they proceed towards the center before eventually disappearing at $r = 0$. All of the compact stellar spheres in the current scenario assures Buchdahl requirement, and
the highest value for surface red-shift is always $z_{s} \leq 5.211$, demonstrating the stability of our system \cite{Ivanov}.
TABLE $(\ref{tab2})$ shows the statistical data of $z_{s}$ and $\frac{M}{R}$ for the observed seven stars near the center, which also indicate that at the center these characteristics are absolutely finite.

\begin{table}[ht]
\caption{Data of redshift component and mass radius ratio.}
\centering
\centering
\begin{tabular}{|p{4.2cm}|p{2.9cm}| p{2.9cm}| | p{2.9cm}|}
\hline
\hline
\begin{center}
  Objects
\end{center}
& ~~~\begin{center}
     $z_{s}$
    \end{center} & ~~~ \begin{center}
    $\frac{M}{R}$
    \end{center}  & ~~~ \begin{center}
    $u(r)$
    \end{center} \\
    \hline

~~~~~~~$RXJ 1856-37 (s1)~~$  &~~~~~~~~$0.282535$ &~~~~~~~$0.22348$ &~~~~~~~$0.196029$ \\\hline
~~~~~~~~~$Her X-1 (s2)$~~  &~~~~~~~~$0.202954$ &~~~~~~~$0.169498$ &~~~~~~~$0.154481$ \\\hline
~~~~~~~~~$LMC X-4 (s3)$~~  &~~~~~~~~$0.202954$ &~~~~~~~$0.185813$ &~~~~~~~$0.154481$ \\
\hline
\end{tabular}
\label{tab2}
\end{table}
\subsection{Casuality Condition}
The analysis of physically compatible models depends critically on the stability of celestial structure. Seeing such celestial objects with stable behaviour against external disturbances is more exciting. In order to explore evolution of celestial bodies, the phenomenon of stability has attracted a lot of attention. Here, we use speed of sound $v_{s}^{2}$ techniques contingent on Herrera's exploded idea \cite{Herrera}, to analyse the stability of our selected stars. The sound speed, denoted by $v_{s}^{2}=dp/d\rho$ must, by the causality requirement, be in the range [0, 1] everywhere inside the stars for a actually steady celestial body. For the anisotropic fluid, sound speed must satisfy the conditions i.e. $0\leq v_{sr}^{2}\leq1$ and $0\leq v_{st}^{2}\leq 1$, where $v_{sr}$ and $v_{st}$ stand for the radial and transverse elements of the sound speed, respectively.\\
In order to investigate likely steady/unsteady configurations of compact stars, Herrera  \cite{Herrera}  developed the notion of cracking using a distinctive method. The difference between the speed of sound traveling  is used to calculate the possibly steady/unsteady zones. For theoretically stable zones, a positive difference within the elements of the sound speed is required, whereas for unsteady parts, their distinction does not assure the inequality $0\leq\mid v_{st}^{2}-v_{sr}^{2}\mid\leq1$. Figs. $(\ref{14})$-$(\ref{16})$ display the stability  of our suggested stars for various $\mathcal{B}$ values, which depicts $0\leq v_{sr}^{2}\leq1$, $0\leq v_{st}^{2}\leq1$ and $0\leq\mid v_{st}^{2}-v_{sr}^{2}\mid\leq1$.

\begin{figure}[h!]
\begin{tabular}{cccc}
\epsfig{file=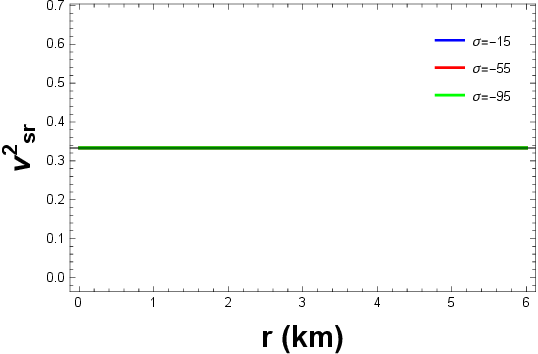, width=.30\linewidth} &
\epsfig{file=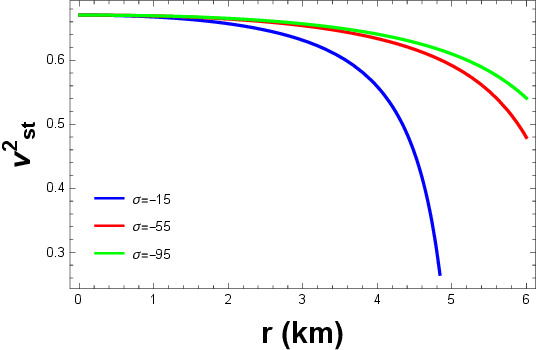, width=.30\linewidth} &
\epsfig{file=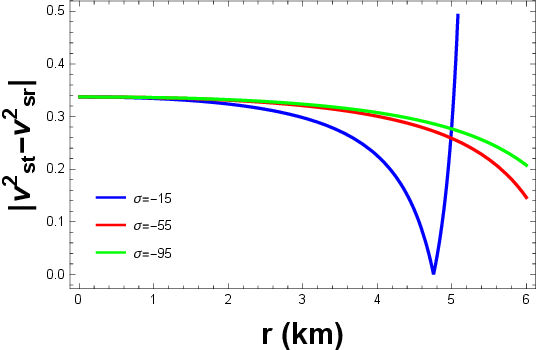, width=.30\linewidth} &
\end{tabular}
\caption{{Deviation of sound speed against $r$ conform to MIT $BM$ EOS for $s1$}.}
\label{Fig.14}
\end{figure}

\begin{figure}[h!]
\begin{tabular}{cccc}
\epsfig{file=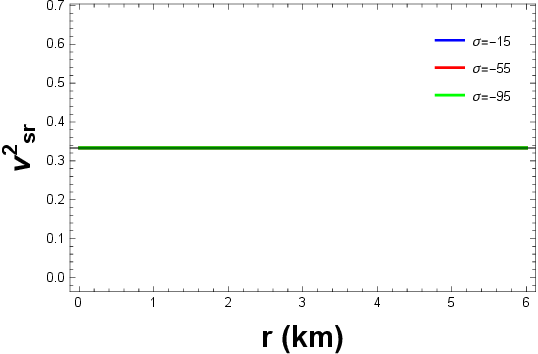, width=.30\linewidth} &
\epsfig{file=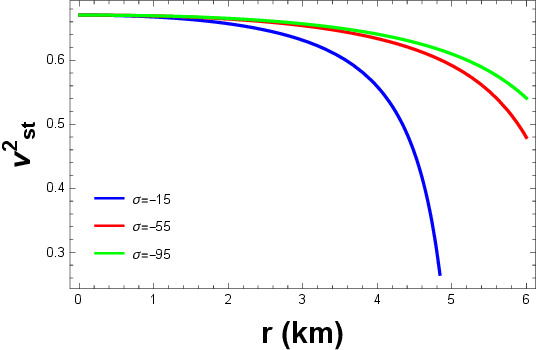, width=.30\linewidth} &
\epsfig{file=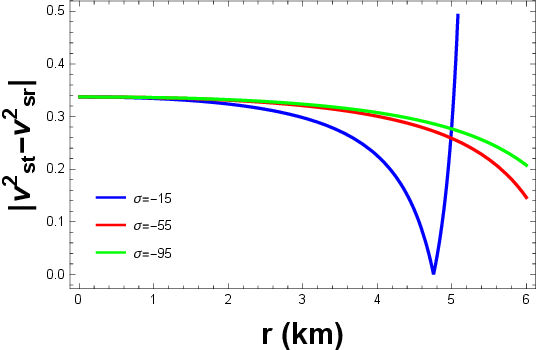, width=.30\linewidth} &
\end{tabular}
\caption{{Deviation of sound speed against $r$ conform to MIT $BM$ EOS for $s2$}.}
\label{Fig.15}
\end{figure}

\begin{figure}[h!]
\begin{tabular}{cccc}
\epsfig{file=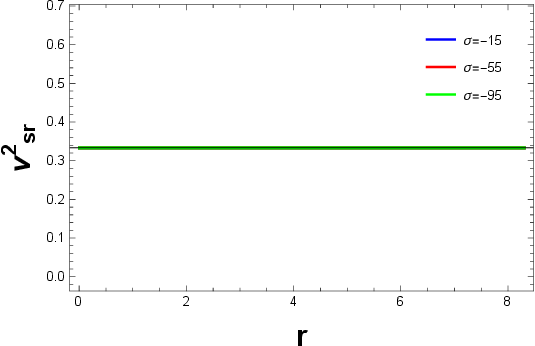, width=.30\linewidth} &
\epsfig{file=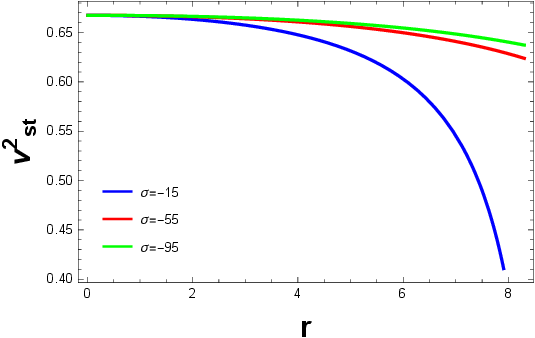, width=.30\linewidth} &
\epsfig{file=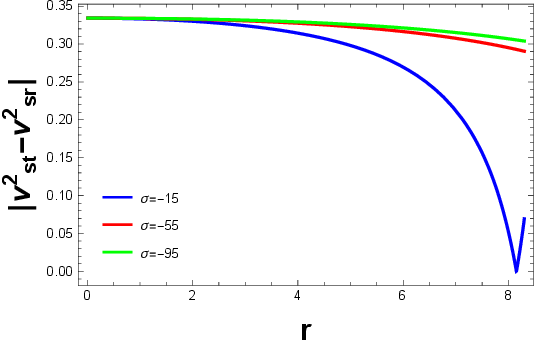, width=.30\linewidth} &
\end{tabular}
\caption{{Deviation of sound speed against $r$ conform to MIT $BM$ EOS for $s3$}.}
\label{Fig.16}
\end{figure}

\subsection{Tolman-Oppenheimer-Volkoff Equation for Modified $f(R,T)$ Gravity}
The Tolman-Oppenheimer-Volkoff $(TOV)$ equation is well recognized for its importance in the study of compact object's equilibrium. Initially, Tolman \cite{Tolman} and subsequently Oppenheimer and Volkoff \cite{Opp} showed the equilibrium condition of the compact stars for the stable $f(R,T)$ model.
\begin{equation}\label{47}
\frac{2}{r}\Delta-\frac{dp_{r}}{dr}-\frac{\sigma'}{2}(\rho+pr)+\frac{2\phi}{3(8\pi-\phi)}\frac{d}{dr}(3\rho-p_{r}-2p_{t})=0.
\end{equation}
According to them, the union of these forces, namely anisotropic force $(F_{a})$, hydrostatic force $(F_{h})$,
gravitational force $(F_{g})$ and an additional force $(F_{frt})$, due to modification of $f(R,T)$ should be zero.
\begin{equation}
F_{a}+F_{h}+F_{g}+F_{frt}=0.
\end{equation}

Fig. $(\ref{17})$ show the graphical depiction of all forces, demonstrating that our system satisfies all equilibrium requirements and demonstrating the feasibility of our selected model.

\begin{figure}[h!]
\begin{tabular}{cccc}
\epsfig{file=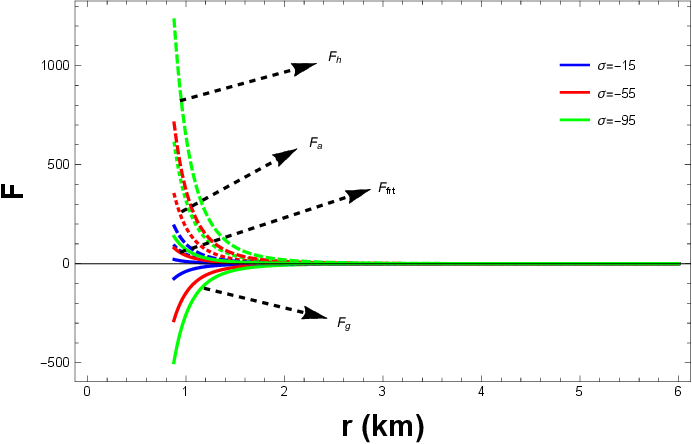,width=0.30\linewidth} &
\epsfig{file=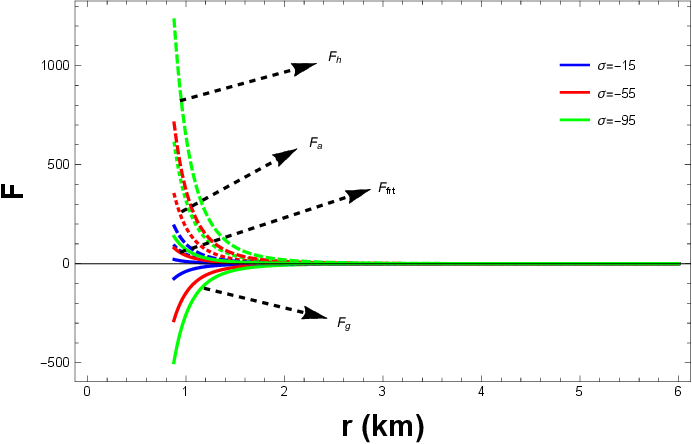,width=0.30\linewidth} &
\epsfig{file=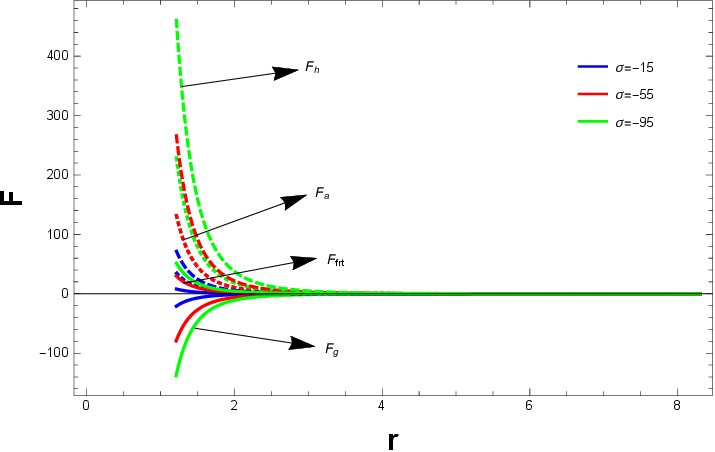,width=0.30\linewidth} &
\end{tabular}
\caption{{Behaviour of TOV equation for $s1$}, $s2$ and $s3$.}
\label{Fig.17}
\end{figure}

\section{Concluding Remarks}
The interaction of the geometry and matter components in alternative theories is essential to narrate interesting phenomena of universe's current accelerating expansion. The modified $f(R,T)$ gravity in this situation offers an interesting  aspect without any enigmatic energy component. In this theory, the junction of geometry and matter components generate the non-zero covariant derivative of $EMT$, which is a prominent characteristic to explore the properties of gravity at the quantum level and investigates the influence of non-geodesic motion of test particles.
\begin{itemize}
\item In the context of $f(R,T)$ theory, this research investigates the impact of anisotropy and $\mathcal{B}$ on the physical characteristics of three specific compact objects. We took into account that the fluid dispersion inside of the stellar structure is simulated using the MIT $BM$ $EOS$ in order to investigate the impact of matter-curvature-coupling existing in a realistic and feasible model in the formation of anisotropic $QS$ applicants.

\item We used Tolman V metric potentials for the realistic modeling of compact stars, in which the unknown parameters $Y$, $b$ and $A$ are determined via a smooth junction between the inner and outer configurations of anisotropic star models. The values of $Y$, $b$ and $A$  are determined for $\omega=-15, -55, -95$ and $\phi=-1.1$ by utilizing the masses as well as radii of suggested compact star models given in TABLE $(\ref{Table 1})$.

\item  Energy density and pressure constituents have been visibly portrayed in our present work in FIGs. $(\ref{Fig.1})$-$(\ref{Fig.3})$, with respect to a specific value of $\mathcal{B}$. It can be clearly seen that these physical parameters attain finite value at the boundary while it becomes infinite at the center. This discovery supports the remarkable density at the star's core.

 \item The pictorial examination in FIGs. $(\ref{Fig.4})$-$(\ref{Fig.6})$, shown that the derivatives of pressure and density are negative, proving that they are monotone decreasing functions of $r$.

 \item From FIG. $(\ref{Fig.7})$, it is shown that anisotropy for our model under consideration is positive, which shows that the behaviour of anisotropic force is repulsive.

 \item It can be seen from FIGs. $(\ref{Fig.8})$-$(\ref{Fig.10})$, that the solution under examination assures energy conditions, illustrating the presence of regular and realistic matter within the stellar object.

\item The computed mass is shown in FIGs. $(\ref{Fig.11})$-$(\ref{Fig.13})$ to be remarkably close to the statistical data, suggesting that our mass function is conventional at the centre. Furthermore, the compactness factor and redshift function are in their permitted limits.

\item The radial and tangential sound speed is frequently in the stability criteria i.e. $0\leq |v^2_{t}-v^2_{r}|\leq1$ as shown by FIGs. $(\ref{Fig.14})$-$(\ref{Fig.16})$,confirming the dynamical stability of our system.

\item FIG. $(\ref{Fig.17})$ shows that all of the forces are in balance, indicating that our system is steady.
   \end{itemize}
The above outcomes are also represented in the following Table $(\ref{tab3})$.\\
\begin{table}[ht]
\caption{Summary of calculated results using observed values of considered stars.}
\centering
\centering
\begin{tabular}{|p{3cm}|p{4.1cm}|p{4.1cm}| p{4.1cm}| }
\hline
\hline

\begin{center}
     $Expression$
    \end{center} & ~~~\begin{center}
     $RXJ1856-37~~(S1)$
    \end{center} & ~~~\begin{center}
     $Her X-1~~(S2)$
    \end{center} & ~~~ \begin{center}
    $LMC X-4~~(S3)$
    \end{center} \\
    \hline

$~~~~~~~~~~~\rho$ &~~~~~~~~~~$>0$, fulfilled &~~~~~~~~~~$>0$, fulfilled &~~~~~~~~~$>0$, fulfilled \\\hline
$~~~~~~~~~~~p_{r}$  &~~~~~~~~~~$>0$, fulfilled &~~~~~~~~~~$>0$, fulfilled &~~~~~~~~~$>0$, fulfilled  \\\hline
$~~~~~~~~~~~p_{t}$ &~~~~~~~~~~$>0$, fulfilled &~~~~~~~~~~$>0$, fulfilled &~~~~~~~~~$>0$, fulfilled  \\\hline
$~~~~~~~~~~\frac{d\rho}{dr}$  &~~~~~~~~~~$<0$, fulfilled &~~~~~~~~~~$<0$, fulfilled &~~~~~~~~~$<0$, fulfilled  \\\hline
$~~~~~~~~~~\frac{dp_{r}}{dr}$ &~~~~~~~~~~$<0$, fulfilled &~~~~~~~~~~$<0$, fulfilled &~~~~~~~~~$<0$, fulfilled  \\\hline
$~~~~~~~~~~\frac{dp_{t}}{dr}$ &~~~~~~~~~ $<0$, fulfilled &~~~~~~~~~~$<0$, fulfilled &~~~~~~~~~$<0$, fulfilled \\\hline
$~~~~~~~~~~~\Delta$ &~~~~~~~~~~$>0$, fulfilled &~~~~~~~~~~$>0$, fulfilled &~~~~~~~~~$>0$, fulfilled \\\hline
$~~~~\rho+p_{r}-C$ &~~~~~~~~~~$>0$, fulfilled &~~~~~~~~~~$>0$, fulfilled &~~~~~~~~~$>0$, fulfilled  \\\hline
$~~~~\rho+p_{t}-C$ &~~~~~~~~~~$>0$, fulfilled &~~~~~~~~~~$>0$, fulfilled &~~~~~~~~~$>0$, fulfilled  \\\hline
$~~~~\rho-p_{r}-C$ &~~~~~~~~~~$>0$, fulfilled &~~~~~~~~~~$>0$, fulfilled &~~~~~~~~~$>0$, fulfilled  \\\hline
$~~~~\rho-p_{t}-C$ &~~~~~~~~~~$>0$, fulfilled &~~~~~~~~~~$>0$, fulfilled &~~~~~~~~~$>0$, fulfilled  \\\hline
$~~~\rho+p_{r}+2p_{t}-C$ &~~~~~~~~~~$>0$, fulfilled &~~~~~~~~~~$>0$, fulfilled &~~~~~~~~~$>0$, fulfilled  \\\hline
$~~~~~~~\rho-C$ &~~~~~~~~~~$>0$, fulfilled &~~~~~~~~~~$>0$, fulfilled &~~~~~~~~~$>0$, fulfilled  \\\hline
$~~~~~~~~~M(r)$ &~~~~~~~~~ $>0$, fulfilled &~~~~~~~~~~$>0$, fulfilled &~~~~~~~~~$>0$, fulfilled \\\hline
$~~~~~~~~~~u(r)$ &~~~$0<u(r)<\frac{8}{9}$, fulfilled &~~~$0<u(r)<\frac{8}{9}$, satisfied &~~~$0<u(r)<\frac{8}{9}$, fulfilled \\\hline
$~~~~~~~~~~~z_{s}$ &~~~~$0<z_{s}<5$, fulfilled &~~~~$0<z_{s}<5$, fulfilled &~~~~$0<z_{s}<5$, fulfilled \\\hline
$~~~~~~~~~~~v^{2}_{sr}$ &~~~~$0<v^{2}_{sr}<1$, fulfilled &~~~~$0<v^{2}_{sr}<1$, fulfilled &~~~~$0<v^{2}_{sr}<1$, fulfilled \\\hline
$~~~~~~~~~~~v^{2}_{st}$ &~~~~$0<v^{2}_{st}<1$, fulfilled &~~~~$0<v^{2}_{st}<1$, fulfilled &~~~~$0<v^{2}_{st}<1$, fulfilled \\\hline
$~~~~~~~~v^{2}_{st}-v^{2}_{sr}$ &$-1<v^{2}_{st}-v^{2}_{sr}<1$, fulfilled &$-1<v^{2}_{st}-v^{2}_{sr}<1$, fulfilled &$-1<v^{2}_{st}-v^{2}_{sr}<1$, fulfilled \\\hline
$~~~~F_{a},F_{h},F_{g},F_{frt}$ &~~~~~~~~~~~~~~Balanced &~~~~~~~~~~~~~~Balanced &~~~~~~~~~~~~~Balanced  \\
\hline
\end{tabular}
\label{tab3}
\end{table}
 
At last, it is worth mentioning that every anisotropic $SSS$ demonstrated in the present work approving obtained well-behaved stars by utilizing the Tolman V potentials. It is important to highlight out that even in the presence of higher-curvature terms in $f(R,T)$  gravity, stellar objects in the context of the MIT $BM$ $EOS$ exhibit uniform and stable patterns, demonstrating the validity of our suggested $f(R,T)$ model (\ref{24}).

\section*{Conflict of Interest}
\hskip\parindent
\small
The authors declare that they have no conflict of interest.

\section*{Contributions}
\hskip\parindent
\small
 We declare that all the authors have same contributions to this paper.

\section*{Data Availability Statement}
The authors declare that the data supporting the findings of this study are available within the article.

\section*{Acknowledgement}
Adnan Malik acknowledges the Grant No. YS304023912 to support his Postdoctoral Fellowship at Zhejiang Normal University, China.

\end{document}